\documentclass{article}

\usepackage{PRIMEarxiv}

\usepackage[utf8]{inputenc} 
\usepackage[T1]{fontenc}    
\usepackage{hyperref}       
\usepackage{url}            
\usepackage{booktabs}       
\usepackage{amsfonts}       
\usepackage{nicefrac}       
\usepackage{microtype}      
\usepackage{lipsum}
\usepackage{fancyhdr}       
\usepackage{graphicx}       
\graphicspath{{media/}}     

\usepackage{hyperref}
\hypersetup{
colorlinks = true, 
linkcolor=black, 
citecolor=black, 
urlcolor=black 
}
\usepackage{amsmath}
\usepackage{tikz}
\usetikzlibrary{quantikz2}
\usepackage{caption}
\usepackage{subcaption}
\usepackage{multirow}
\usepackage{xspace}
\usepackage{makecell}
\usepackage{xcolor}
\usepackage{mathtools}
\usepackage{paralist}
\usepackage{subcaption}
\usepackage{pifont}
\usepackage[most,breakable]{tcolorbox}
\usepackage{booktabs}
\usepackage{xspace}
\usepackage{bbding}
\usepackage{multirow}
\usepackage{multicol}
\usepackage{amsthm}
\usepackage{caption}
\usepackage{xcolor,colortbl}
\usepackage{subcaption}
\usepackage{tabularx}
\usepackage{siunitx}
\usepackage{stfloats}
\usepackage{nicefrac}
\usepackage[inline]{enumitem}
\definecolor{lightgreen}{rgb}{0.56, 0.93, 0.56}
\newcommand{\ourApproach}{\ensuremath{\textsc{QUELL}}\xspace}
\newcommand{\dario}{\ensuremath{\textsc{DARIO}}\xspace}
\newcommand{\mlApproach}{\ensuremath{\textsc{DARIO}_{\textsc{pred}}}\xspace}
\newcommand{\mlApproachSVM}{\ensuremath{\textsc{DARIO}_{\textsc{pred}}^{\textsc{SVM}}}\xspace}
\newcommand{\mlApproachRT}{\ensuremath{\textsc{DARIO}_{\textsc{pred}}^{\textsc{RT}}}\xspace}
\newcommand{\Elevate}{\textsc{Elevate}\xspace}
\newcommand{\svector}[1]{\ensuremath{\boldsymbol{S}_{#1}}\xspace}
\newcommand{\svalue}[2]{\ensuremath{s}_{#1#2}\xspace}

\newcommand{\feaAngle}[2]{\ensuremath{\theta_{#1}^{#2}}\xspace}
\newcommand{\feature}[1]{\ensuremath{\mathit{F}_{#1}}\xspace}
\newcommand{\timeWin}[1]{\ensuremath{\mathit{tw}_{#1}}\xspace}
\newcommand{\target}[1]{\ensuremath{t_{#1}}\xspace}
\newcommand{\thetaVector}[1]{\ensuremath{\boldsymbol{\theta}_{#1}}\xspace}
\newcommand{\featureSetName}{\ensuremath{\mathit{FS}}\xspace}
\newcommand{\featureSet}[1]{\ensuremath{\featureSetName_{\mathit{#1}}}\xspace}
\newcommand{\amse}{\ensuremath{\mathit{AMSE}}\xspace}
\newcommand{\mse}{\ensuremath{\mathit{MSE}}\xspace}
\newcommand{\expDayName}{\ensuremath{\mathit{ExpDay}}\xspace}
\newcommand{\expDay}[1]{\ensuremath{\expDayName_\mathit{#1}}\xspace}

\newcommand{\Atwelve}{\ensuremath{\hat{A}}\textsubscript{12}\xspace}

\newcommand{\CZ}{\ensuremath{\mathit{CZ}}\xspace}
\newcommand{\CX}{\ensuremath{\mathit{CX}}\xspace}
\newcommand{\RX}{\ensuremath{\mathit{RX}}\xspace}
\newcommand{\RY}{\ensuremath{\mathit{RY}}\xspace}
\newcommand{\RZ}{\ensuremath{\mathit{RZ}}\xspace}
\newcommand{\CNOTgate}{\ensuremath{\mathit{CNOT}}\xspace}
\newcommand{\CNOTreservoir}{\ensuremath{\mathtt{CNOT}}\xspace}

\newcommand{\DHE}{\ensuremath{\mathtt{DHE}}\xspace}
\newcommand{\RHE}{\ensuremath{\mathtt{RHE}}\xspace}
\newcommand{\Rotation}{\ensuremath{\mathtt{Rotation}}\xspace}
\newcommand{\Ising}{\ensuremath{\mathtt{ISING}}\xspace}
\newcommand{\Harr}{\ensuremath{\mathtt{Harr}}\xspace}
\newcommand{\combination}[2]{#1\_\-#2}

\newcommand{\combEncRes}{\combination{{\tt En\-co\-der}}{{\tt Res\-er\-voir}}\xspace}

\newcommand{\AWT}{\ensuremath{\mathtt{AWT}}\xspace}
\newcommand{\predictedValue}{\ensuremath{t^{\mathit{pre}}}\xspace}

\title{Application of Quantum Extreme Learning Machines for QoS Prediction of Elevators' Software in an Industrial Context
}

\author{
  Xinyi Wang\\
  Simula Research Laboratory \\
  University of Oslo \\
  Oslo, Norway\\
  \texttt{xinyi@simula.no} \\
   \And
  Shaukat Ali \\
  Simula Research Laboratory \\
  Oslo Metropolitan University\\
  Oslo, Norway\\
  \texttt{shaukat@simula.no} \\
  \AND
  Aitor Arrieta \\
  Mondragon University \\
  Mondragon, Spain \\
  \texttt{aarrieta@mondragon.edu} \\
  \And
  Paolo Arcaini \\
  National Institute of Informatics \\
  Tokyo, Japan \\
  \texttt{arcaini@nii.ac.jp} \\
  \And
  Maite Arratibel \\
  Orona \\
  San Sebastian, Spain \\
  \texttt{marratibel@orona-group.com} \\
}

\begin{document}
\maketitle

\begin{abstract}
Quantum Extreme Learning Machine (QELM) is an emerging technique that utilizes quantum dynamics and an easy-training strategy to solve problems such as classification and regression efficiently. Although QELM has many potential benefits, its real-world applications remain limited. To this end, we present QELM's industrial application in the context of elevators, by proposing an approach called \ourApproach. In \ourApproach, we use QELM for the waiting time prediction related to the scheduling software of elevators, with applications for software regression testing, elevator digital twins, and real-time performance prediction. The scheduling software has been implemented by our industrial partner Orona, a globally recognized leader in elevator technology. We demonstrate that \ourApproach can efficiently predict waiting times, with prediction quality significantly better than that of classical ML models employed in a state-of-the-practice approach. Moreover, we show that the prediction quality of \ourApproach does not degrade when using fewer features. Based on our industrial application, we further provide insights into using QELM in other applications in Orona, and discuss how QELM could be applied to other industrial applications.
\end{abstract}



\maketitle

\section{Introduction}\label{sec:introduction}
Orona is a Spanish company well-renowned for building elevators~\cite{oronaWebsite}, i.e., vertical transportation systems responsible for safely, comfortably, and efficiently transporting passengers across different floors. At the core of elevators, there is the {\it dispatcher}, a scheduling software that assigns an elevator to each call by maintaining an acceptable {\it quality of service} (QoS). A commonly used metric for measuring the QoS of the elevators is the {\it Average Waiting Time} (\AWT), i.e., the average time passengers have to wait, from when they press the call button until the time the elevator arrives. \AWT is widely used to test dispatchers under various building configurations to identify quality issues~\cite{ayerdi2022performance,ValleICSE}, followed by a proper dispatcher configuration to achieve acceptable performance.

The dispatcher software undergoes evolution like any other software system, requiring regression testing. In Orona, a regression test oracle is employed during design time in both Software in the Loop (SiL) and Hardware in the Loop (HiL) settings. To reduce testing costs, such regression test oracle was proposed to be replaced by a machine learning(ML)-based test oracle~\cite{AitorPaper}.

Currently, as the DevOps paradigm emerges in the context of Cyber-Physical Systems (CPSs), Orona is interested in deploying such oracles also in operation to see whether the QoS of the elevators meets acceptable values. However, the application of existing ML-based oracles is not straightforward. Indeed, during design time, several features are available. For example, the passengers' weight is a feature that an ML-based test oracle can use at design time to determine whether the \AWT of passengers is within an acceptable threshold in a given time window. However, some of these features, such as the passengers' weight itself, are not easy to measure at operation time; hence, they cannot be used as features for ML models to be used at operation time. Moreover, elevators must be configured according to the constraints imposed by the building where they are installed. For example, in some elevators, it is possible to extract the exact distance traveled by each passenger 
In contrast, in other elevators, it is not possible. As a result, different number of features are available for different elevator installations.



Hence, in Orona, there is an increasing need for ML-based test oracles that can be trained on a variable number of features, ranging from a small number to a large number of features, depending on the installation configurations. Moreover, there is interest in using ML models at the operation time to support analyses such as the predictions of various QoS, runtime verification and monitoring, and advanced analyses with digital twins of elevators~\cite{DTEvolution}. 

To cater to the above needs of Orona, we explore the use of Quantum Extreme Learning Machine (QELM)~\cite{QELMOpportun}, by proposing a QELM-based approach called {\it QUantum Extreme Learning eLevator} (\ourApproach). 
QELM is a quantum machine learning technique that uses quantum dynamics of quantum reservoirs to enable a simple machine learning model (e.g., a linear regression model) to be efficiently trained with a limited number of features, but still provide good prediction quality. The ability of the quantum reservoir to map input data with limited dimensions into a higher-dimensional quantum space contributes to this efficiency. 

To assess the effectiveness and efficiency of \ourApproach, we used operational data from four days of operation of a real elevator installation. We compare it with the existing ML-based regression models employed in the same industrial context for predicting \AWT. The results showed that \ourApproach can significantly outperform the classical ML-based regression models for the prediction task, thereby demonstrating QELM's potential benefits in an industrial context. Based on our results, we discuss the application of \ourApproach to various contexts at Orona, followed by a discussion on the potential applications in other industrial contexts valuable for practitioners. Finally, we present open research questions related to QELM applications in software engineering that are valuable for software engineering researchers.


\section{Background} \label{sec:background}
We provide quantum computing basics and background on QELM.

\smallskip
\noindent{\bf Quantum Computing Basics.}
Quantum computers compute with {\it quantum bits} (i.e., {\it qubits}). Unlike a classical bit taking values in $\{0, 1\}$, a qubit can be in {\it superposition}, i.e., in states 0 and 1 simultaneously, with specific probabilities of collapsing into either 0 or 1 upon measurement. A {\it quantum state} can be represented as a {\it state vector}, e.g., a one-qubit state can be represented as $\ket{\psi}=\alpha_0\ket{0}+\alpha_1\ket{1}$, where $\alpha_0$ and $\alpha_1$ are the {\it amplitude} of the quantum state. $|\alpha_0|^2$ and $|\alpha_1|^2$ represent the probability of being in state $\ket{0}$ and $\ket{1}$ when observed. Similarly, a $D$-qubit state vector can be represented as:
\begin{equation*}
\textstyle \ket{\psi} = \left(
\begin{array}{c}
\alpha_{0}\\
\vdots \\
\alpha_{d}
\end{array}
\right) = \alpha_0\ket{0}+
\ldots+\alpha_{d}\ket{d}\quad\quad\text{with} \; d=2^D
\end{equation*}
$\ket{\psi}$ is a normalized $d$-dimensional vertical vector, i.e., $\sum_{i=0}^{d}|\alpha_i|=1$.

Quantum computers are currently programmed with {\it quantum circuits}. A quantum circuit consists of a series of {\it quantum gates} (e.g., a Hadamard gate that places a qubit into a superposition) performing computations on qubits, i.e., evolving the quantum state over time, guided by the principles of quantum mechanics implemented in the quantum gates. This time evolution process is defined by a {\it Hamiltonian}\footnote{A Hamiltonian defines the total energy of a quantum system. It is specified as a {\it hermitian matrix}.} $H$ applied to the quantum circuit. We represent the process over a period of time $\Delta t$ as $\ket{\psi} = \mathit{exp}(-iH\Delta t\ket{\psi_0})$, where the quantum state evolves from $\ket{\psi_0}$ to $\ket{\psi}$. In this equation, $\mathit{exp}(-iH\Delta t)$ is a quantum gate, i.e., a unitary operator. The operator can also be represented as $U$ and the computation process as $\ket{\psi} = U\ket{\psi_0}$. Table~\ref{table:gatetype} describes the gates used in \ourApproach.
\begin{table}[!tb]
\caption{Descriptions of quantum gates used in \ourApproach}
\label{table:gatetype}
\footnotesize
\resizebox{\columnwidth}{!}{
\begin{tabular}
{m{0.19\columnwidth}|m{0.8\columnwidth}}
\toprule
\textbf{Gate} & \textbf{Description} \\
\midrule
\textit{Pauli-Z} (\textit{Z}/$\sigma_z$) & It rotates a qubit around the $z$-axis with $\pi$ radians. \\
\hline
\textit{Pauli-X} (\textit{X}/$\sigma_x$) & It rotates a qubit around the $x$-axis with $\pi$ radians. \\
\hline
\textit{Pauli-Y} (\textit{Y}/$\sigma_y$) & It rotates a qubit around the $y$-axis with $\pi$ radians. \\
\hline
$\RX(\theta)$ & It rotates a qubit around the $x$-axis with $\theta$ radians. \\
\hline
$\RY(\theta)$ & It rotates a qubit around the $y$-axis with $\theta$ radians.\\
\hline
$\RZ(\theta)$ & It rotates a qubit around the $z$-axis with $\theta$ radians. \\
\hline
\textit{Controlled-NOT (\CNOTgate/\CX)} & A two-qubit gate with a control and a target qubit. If the control qubit is $\ket{1}$, the target qubit rotates around the x-axis with $\pi$ radians. \\
\hline
\textit{Controlled-Z (\CZ)} & A two-qubit gate with a control and a target qubit. If the control qubit is $\ket{1}$, the target qubit rotates around the x-axis with $\pi$ radians. \\
\bottomrule
\end{tabular}
}
\end{table}

At the end of circuit execution, the final quantum state is measured with a set of {\it observables} (unitary operators) as the output of the circuit. {\it Pauli operators} (i.e., $\textit{X}, \textit{Y}, \textit{Z}$) are usually used to obtain partial information of $x$, $y$, or $z$ basis from each qubit. The measured result of each qubit is determined by the expectation values. For example, if the $j$-th qubit in $z$ basis is measured, the observed expectation value of the operator $Z$ is represented as $\left\langle Z^j \right\rangle$.

\smallskip
\noindent{\bf Quantum Extreme Learning Machine.}
{\it Extreme learning machine} (ELM) is a feedforward neural network that uses fixed, nonlinear dynamics to extract information from inputs efficiently. It has neurons in the hidden layers and an output function~\cite{huang2006extreme}. The parameters of hidden layers are fixed and assigned randomly instead of being optimized during training, as in classic neural networks. ELM only needs to train the weights of the output function, which is usually a linear regression algorithm to compute the output weights.

\begin{figure}[!tb]
    \centering
    \includegraphics[width=0.7\columnwidth]{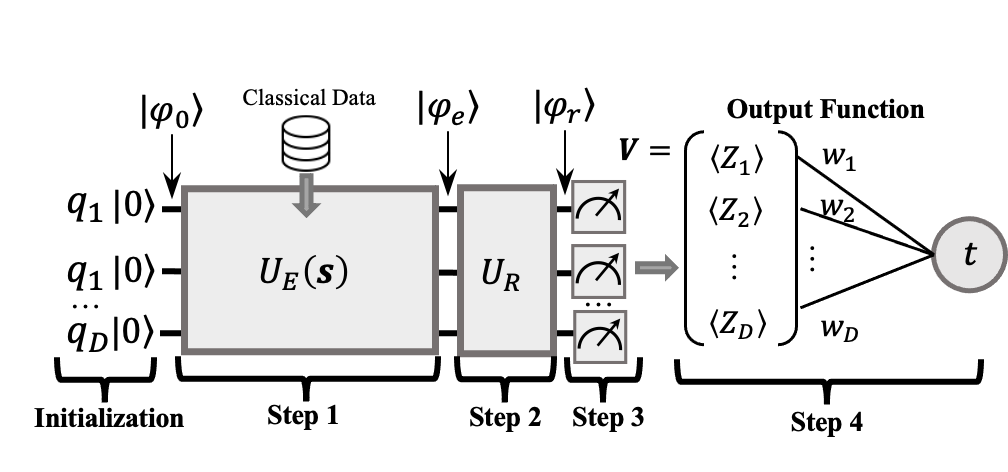}
    \caption{Implementation of QELM}
    \label{fig:background}
\end{figure}

A {\it quantum extreme learning machine} (QELM) is the quantum counterpart of an ELM. A key advantage of QELMs is that they can use the complex dynamics of a quantum system evolution process in an exponentially large quantum space. They enable intricate feature mapping for classical data inputs. The quantum circuit providing high dynamics is called \emph{quantum reservoir}. The rich dynamics allow the use of far fewer resources than classical ELM, which instead requires extensive hidden layers~\cite{fujii2017harnessing,QELMOpportun}. The QELM's implementation has four steps (see Fig.~\ref{fig:background}).

\noindent\textbf{Step 1: }During the training process, a classical input vector $\boldsymbol{s}$ is first given as input to a {\it quantum encoder} circuit to be transformed into quantum states. The quantum encoder is a set of gates containing parameterized quantum gates. Suppose the quantum circuit is initialized as $\ket{\psi_0}$. The quantum encoder with the classical input is applied on the circuit, obtaining the state $\ket{\psi_e}=U_E(\boldsymbol{s})\ket{\psi_0}$, where $U_E$ represents the unitary of the encoder.

\noindent\textbf{Step 2: }Then, the reservoir is applied to the quantum system. The output state of the encoder goes into a quantum reservoir circuit. The parameters of the quantum reservoir circuits are fixed and randomly assigned. $U_R$ is the unitary of the reservoir. As output, it produces the state $\ket{\psi_r}=U_RU_E(\boldsymbol{s})\ket{\psi_0}$.

\noindent\textbf{Step 3: }Then, a set of observables are applied on the final state $\ket{\psi_r}$ to obtain output information from the quantum circuit. Suppose that there are $D$ qubits and the Pauli $Z$ observable is applied on each qubit to extract information in $z$ basis. The output vector of measured values is $\boldsymbol{V}=[\left\langle Z^1 \right\rangle,$ 
$\ldots,$ $\left\langle Z^D \right\rangle]$.

\noindent\textbf{Step 4: } Finally, as a non-linear function is used in the reservoir, linear regression is used to train the weights of the output function, mapping the observed values $\boldsymbol{V}$ to the target values $t$. The linear regression model tries to find the closest predicted target value \predictedValue:
\begin{equation*}
\textstyle \predictedValue = \sum_{k=1}^M w_k\left\langle Z^k \right\rangle
\end{equation*}
$w_i$ are the weights of the output function. The training minimizes the deviation of the predicted value \predictedValue from the target value $t$.



\section{Industrial Context} \label{sec:industrialcontext}

With over 250,000 elevator installations worldwide, Orona is one of the largest elevator companies in Europe. A system of elevators aims to transport passengers between floors of a building as safely as possible while minimizing the time they need to wait for the elevator. One way to do so is by serving the calls quickly and taking the passengers to their destination in the minimum time possible. There are different ways to accomplish that. For example, faster engines can move elevators faster and, so, serve the calls quicker. Another way this paper considers is scheduling the elevators as optimally as possible. The elevator {\it dispatching algorithm} does this.

The dispatching algorithm is the key component responsible for optimally assigning an elevator to each call. This decision is made by considering inputs as the current position of each elevator, the moving direction, and the number of stops it has already taken. By analyzing these factors, the dispatching algorithm minimizes aspects like the passengers' waiting time or energy consumption.

Orona has an extensive suite of dispatching algorithms. Like any other software system, a dispatching algorithm undergoes regular maintenance and evolution to address hardware obsolescence, adapt to legislative changes, incorporate new functionalities, and resolve bugs. {\it Regression testing} is the predominant technique to ensure a high code quality in response to this evolution. To test such systems, Orona employs simulation-based techniques at different levels. The first level refers to the Software in the Loop (SiL) test level. At this level, the domain-specific simulator \Elevate~\cite{elevateWebsite} is employed. \Elevate takes as inputs a {\it passenger file} and the {\it building configuration file}. The {\it passenger file} encompasses a set of passengers traveling through a building. Each passenger has different attributes, such as the floor they are arriving on, their destination floor, when they arrive, their weight, etc. The {\it building configuration file} contains aspects like the number of elevators, the speed of each of them, the number of each floor, etc. At this level, two types of tests are executed:
\begin{inparaenum}[(i)]
\item {\it short scenario tests} and
\item {\it full-day tests}.
\end{inparaenum}
Short scenario tests focus on examining specific functional properties in an isolated manner. The anticipated results of these tests are achieved through the implementation of assertions, metamorphic testing~\cite{ayerdi2022performance}, or manual methods. During the full-day tests, scenarios resembling a typical full-day or its sub-scenarios in the life-cycle of the elevator system are carried out. The anticipated results of these tests are linked to specific Quality of Service (QoS) values over time, including the \AWT, which are obtained by re-running the test using a different algorithm or an older version. After the SiL phase, during the Hardware-in-the-Loop (HiL) testing phase, similar test processes are iterated, incorporating actual hardware components such as target processors, communication systems, real-time operating systems, and human-machine interfaces. The test cases executed at this level remain consistent with those in earlier stages, but the crucial distinction lies in real-time execution, making it a more costly process.


While classical regression test oracles are affordable in the context of the SiL test level, when considering full-day traffic profiles at the HiL test level, this testing does not scale due to the need to re-execute the previous version (note that at HiL, the test execution is real-time). Moreover, at operation time, it is not possible to re-execute any test. Because of this, recently, the ML-based oracle \mlApproach~\cite{AitorPaper} (i.e., the prediction component of the framework \dario) has been proposed to determine, in an efficient manner, whether a system of elevators gives an adequate QoS.

As explained in Sect.~\ref{sec:introduction}, it is important to use these ML-based oracles in operation to perform additional activities like runtime monitoring, runtime performance prediction, etc. However, \mlApproach is not applicable, as its ML models are trained on many features unavailable at operation time. Indeed, some features, like the passengers' weights, may not be obtainable at runtime. Moreover, the configuration of elevators in different buildings varies, and the number of available features for training ML models varies.

Thus, there is a need to learn an ML model with fewer features than what is possible at the design time while, at the same time, achieving model prediction performance comparable to that obtainable with many features.
To this end, we explore the use of QELM to train an ML model with fewer features for which it is possible to extract the data from the real operation of elevators. We aim to demonstrate that even with few features, we can predict \AWT with comparable prediction performance as with a complete feature set that is possible during design time. We also demonstrate that \mlApproach cannot perform well with fewer features.

\begin{figure}[!tb]
\centering
\includegraphics[width=0.8\columnwidth]{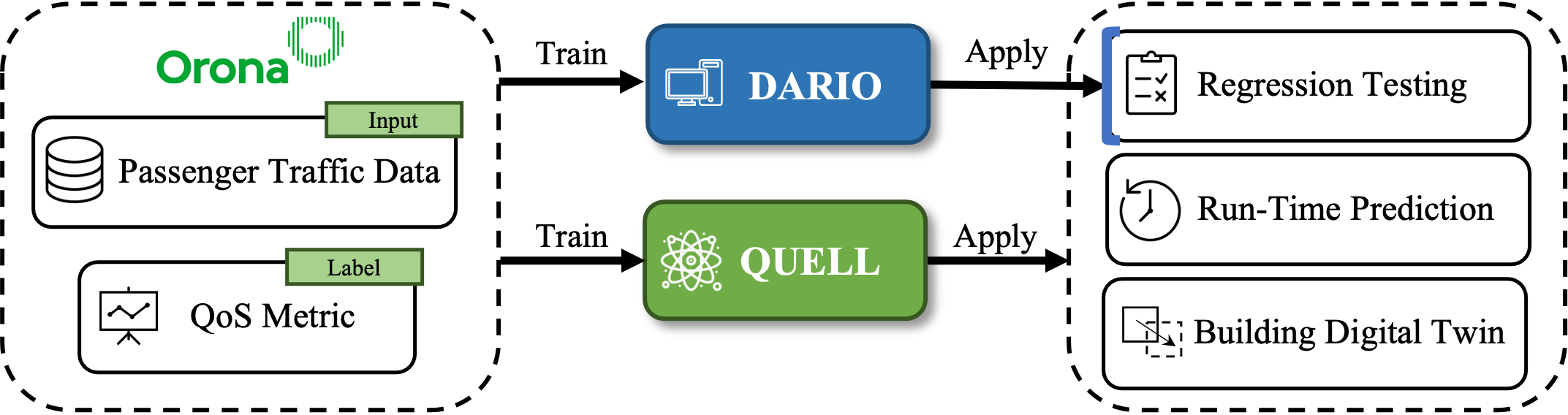}
\caption{Industrial Context of \ourApproach}
\label{fig:indu_con}
\end{figure}

Fig.~\ref{fig:indu_con} shows the industrial context of \ourApproach, where the aim is to support regression testing of the dispatching algorithm.
Regression testing at Orona is currently done with \mlApproach. Another possible application of \ourApproach is supporting real-time analyses by enhancing digital twins of industrial elevators previously built in~\cite{DTEvolution}, and other real-time predictions.

\section{Approach}\label{sec:approach}
In this section, we first present the overview of \ourApproach. Next, we describe two hardware-efficient quantum encoders and the four quantum reservoir circuits we use in this paper.

\subsection{Overview}
\begin{figure}[!tb]
\centering
\includegraphics[width=0.8\columnwidth]{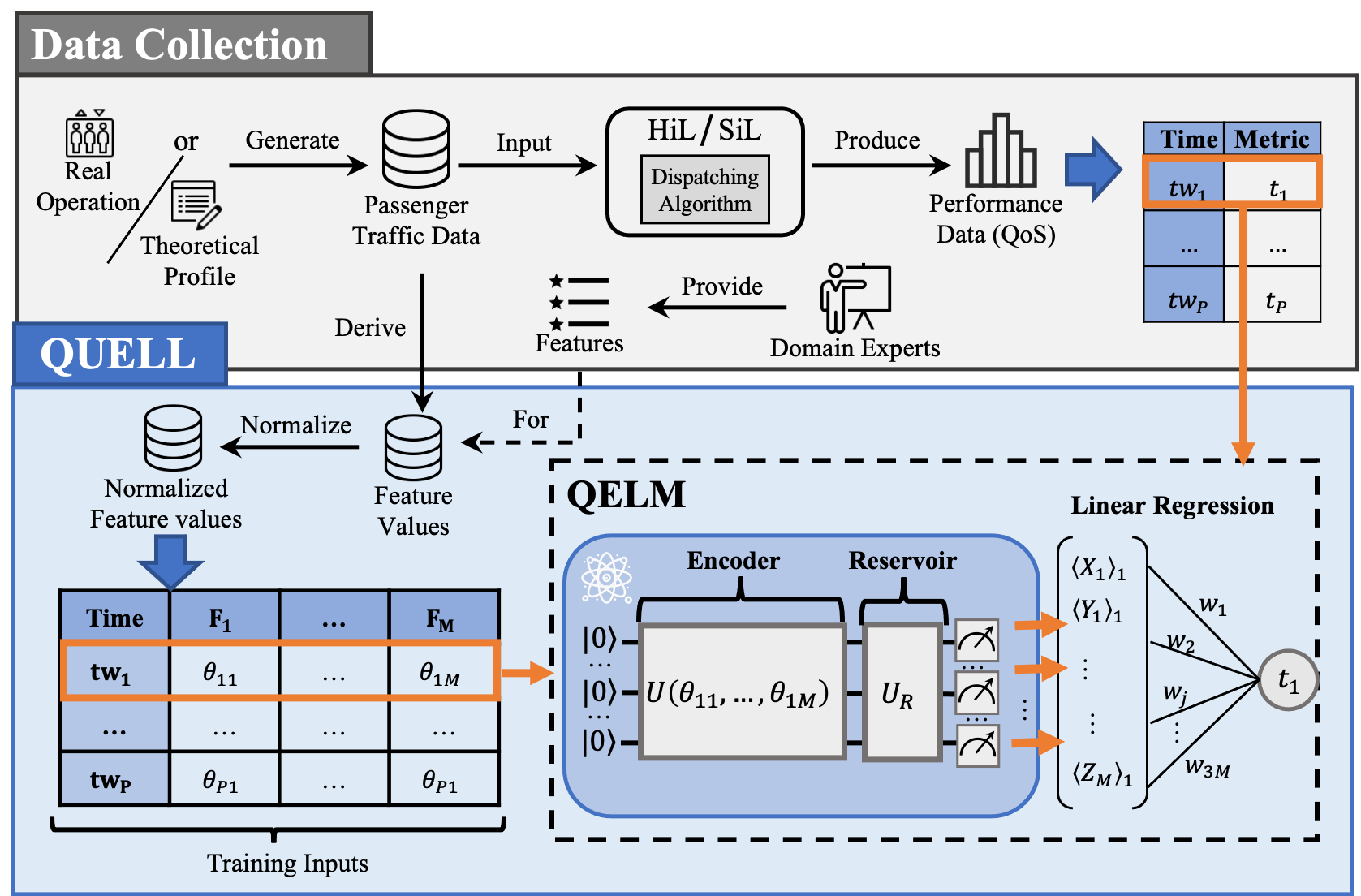}
\caption{Overview of \ourApproach}
\label{fig:overview}
\end{figure}

\ourApproach leverages the capability of QELMs to train the model, which is a hybrid quantum-classical approach. Fig.~\ref{fig:overview} shows the overview of the data collection process and the training process of \ourApproach.
To train the model, \ourApproach requires the previously collected data (i.e., {\it passenger traffic data} from, e.g., the {\it real operation} of industrial elevators) corresponding to domain-specific features developed by {\it domain experts} (elevator designers in our context), and the {\it QoS data} (\AWT in our context) corresponding to QoS metrics that indicate the performance of an elevator. The elevator dispatching algorithm takes input passenger traffic data such as call floor, destination floor, traveling distance, and floor position. During design and development, such data can be simulated or obtained from theoretical traffic profiles and used as input data. During operation, such data is recorded from how real passengers used the elevators. However, during operation time, we can only record limited data, e.g., we can not record individual passengers' weight. In addition, data corresponding to QoS metrics can be extracted from simulations (e.g., in SiL or HiL setup). For SiL, this is achieved by feeding traffic data based on theoretical traffic profiles or real passenger traffic data to the \Elevate simulator and obtaining such QoS values as the model's labels (i.e., {\it training targets}). In the HiL setup, hardware is involved in the simulation to obtain the QoS values. 

Domain experts, i.e., elevator designers from Orona, provide the {\it domain-specific features} for training $\{\feature{1},$ $\ldots,$ $\feature{M}\}$ as shown in Fig.~\ref{fig:overview}. Next, {\it feature values} within each time window \timeWin{j} are obtained. There are two ways to obtain such values. First, they can be obtained via telemetry through the dispatching algorithm in operation. An alternative is to obtain such values from simulations or theoretical traffic profiles, which is common in Orona during design time. However, in the context of this paper, we focused only on the real operational data. Suppose we obtain the {\it feature values} of in total $P$ number of time windows. For each time window, we consider the values as a $M$-dimension input vector $\svector{j}=[\svalue{j}{1},$ $\ldots,$ $\svalue{j}{M}]$ for the model. The corresponding target output value is the functional QoS metric \target{j} obtained in the same time window \timeWin{j}. However, we cannot directly input the raw values into the quantum encoders of QELMs since these values are classical data and must be encoded to be processed by quantum circuits. Such transformation is performed by encoders (see details in Sect.~\ref{subsec:encoders}). The encoders process information by parameterizing the angles of Pauli gates with input data. As a result, we need to consider the periodicity of the angle to avoid encoding several different values in the same quantum state. Arbitrary values will cause inaccurate encoding. Thus, we use min-max normalization on all datasets to normalize all the data, feature by feature, into the range of radians from 0 to $\pi$. The normalized feature data of the $j$-th time window is represented as $\thetaVector{j}=[\feaAngle{j}{1},$ $\ldots, \feaAngle{j}{M}]$. 

Next, we feed the normalized values \thetaVector{j} into a quantum circuit for each time window. Each qubit of a quantum circuit is first initialized with $\ket{0}$ state. Next, we inject \thetaVector{j} into the encoder circuit to transform the classical data into a quantum state. This generated quantum state, which is a hidden state, is the input of the quantum reservoir (see details in Sect.~\ref{subsec:reservoir}). The quantum state evolves inside the quantum reservoir circuit until it is measured. To extract partial information from the x, y, and z axis, we measure the expectation values with all three Pauli operators (i.e., \textit{X}, \textit{Y}, \textit{Z}) on each qubit. Thus, we obtain the output vector of the $j$-th time window as $\boldsymbol{V}_j =$ $[\left\langle X^1 \right\rangle_j,$ $\left\langle Y^1 \right\rangle_j,$ $\left\langle Z^1 \right\rangle_j,$ $\ldots,$ $\left\langle X^M \right\rangle_j,\left\langle Y^M \right\rangle_j, \left\langle Z^M \right\rangle_j]$.

After feeding feature values of all $P$ time windows of the training datasets into the quantum circuit and getting all the measured output vectors, we train the linear regression model to optimize the weights $\boldsymbol{W}=[w_1, w_2, \ldots, w_{3M}]$ of the output function. We try to minimize the difference between the prediction $\predictedValue_j$ and the corresponding training target $t_j$ with a loss function Residual sum of squares (RSS). Thus, we minimize the loss with the equation.
\begin{equation}
\textstyle \mathit{RSS} = \sum_{j=1}^P (\boldsymbol{W}\boldsymbol{V_j}-t_j)^2
\end{equation}

\subsection{Quantum Encoders} \label{subsec:encoders}
We use two quantum encoders to transfer classical data into quantum states before using it in a quantum reservoir circuit to evolve the state. Since hardware restrictions exist on current quantum computers (limited connections among qubits, a limited number of native gates, and the decoherence issue), quantum circuits with high-depth and dense connections are unsuitable. To this end, a class of hardware-efficient ansatz (a parameterized quantum circuit to approximate with a defined sequence of gates to approximate a problem) is proposed~\cite{kandala2017hardware}. This class of circuits has been used to encode classical data for quantum machine learning~\cite{sequeira2022variational,bharti2022noisy}. It typically contains a series of gates, including single-qubit rotation gates followed by some two-qubit entanglement gates. The circuit can be composed of single or multiple layers (i.e., depth) by repeating this set of gates. Also, single or multiple qubits can encode one feature value depending on the size of the problem and the number of available qubits. A higher number of qubits encoding features leads to a higher dimension of the Hilbert space, which adds more dynamics to the circuit. In contrast, a lower number may lack such extensive dynamics but leads to higher efficiency and cost-saving.

The following two encoders exhibit a common gate structure but vary in implementing single qubit rotation gates.

\begin{figure}[!tb]
\centering
\begin{subfigure}[b]{0.49\columnwidth}
\centering
\resizebox{0.75\columnwidth}{!}{ 
\begin{quantikz} [color=black,background color=lightgreen]
&\gate{RX(\feaAngle{j}{1})}& \ctrl{1} & &\gate{Z}&\\
&\gate{RX(\feaAngle{j}{2})}& \gate{Z}&\ctrl{1}& & \\
&\gate{RX(\feaAngle{j}{3})}& &\gate{Z}&\ctrl{-2}& \\
\end{quantikz}
}
\caption{\DHE encoder}
\label{fig:dhe}
\end{subfigure}
\begin{subfigure}[b]{0.49\columnwidth}
\centering
\resizebox{0.75\columnwidth}{!}{ 
\begin{quantikz} [color=black,background color=lightgreen]
&\gate{RZ(\feaAngle{j}{1})}& \ctrl{1} & &\gate{Z}&\\
&\gate{RX(\feaAngle{j}{2})}& \gate{Z}&\ctrl{1}& & \\
&\gate{RY(\feaAngle{j}{3})}& &\gate{Z}&\ctrl{-2}& \\
\end{quantikz}
}
\caption{\RHE encoder}
\label{fig:rhe}
\end{subfigure}
\caption{Two types of quantum encoders used in \ourApproach}
\label{fig:encoders}
\end{figure}
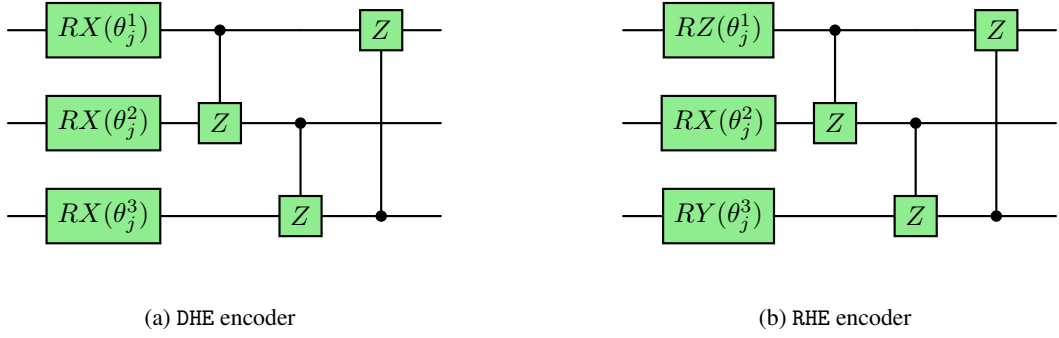

\smallskip
\noindent \textbf{Determined hardware efficient encoder (\DHE):} In each layer, each qubit is first applied with an \RX, and the rotation angles are determined by input values. In our case, for single qubit encoding, each \RX gate is parameterized by one normalized feature value. Specifically, to encode data in the $j$-th time window, i.e., $\thetaVector{j}$, the rotation of \RZ gate on the $k$-th qubit should be determined by $\feaAngle{j}{k}$, which is the normalized value of the $j$-th feature. In the following, \CZ gates are applied with a cyclic entangling structure, where qubits are connected cyclically with successive \CZ gates. An example of one-layer quantum circuit is shown in Fig.~\ref{fig:dhe}, where, given three features selected by domain experts $\{\feature{1},$ $\feature{2},$ $\feature{3}\}$, we encode the values of the $j$-th time window $\thetaVector{j}=[\feaAngle{j}{1},$ $\feaAngle{j}{2},$ $\feaAngle{j}{3}]$ in order on \RX gates for each qubit, followed by the cyclic entangling \CZ gates.

\smallskip
\noindent \textbf{Randomized hardware efficient encoder (\RHE):} This encoder follows the same main structure of the \DHE encoder. The difference is that the \RX gates of \DHE are replaced by randomly selected single qubit rotation gates among \RX, \RY, and \RZ gates. For example, the encoding of the classical data follows the same rule. With the same example illustrated in \DHE, given the same three features in the $j$-th time window and corresponding values $\thetaVector{j}=[\feaAngle{j}{1}, \feaAngle{j}{2}, \feaAngle{j}{3}]$, Fig.~\ref{fig:rhe} shows one possible \RHE circuit of one layer.

\subsection{Quantum Reservoir} \label{subsec:reservoir}
After obtaining the state vectors from the encoders, we use the quantum reservoir circuits to evolve the quantum state by executing the quantum circuit until it is measured. In the following, we introduce four quantum reservoir circuits used in the paper.

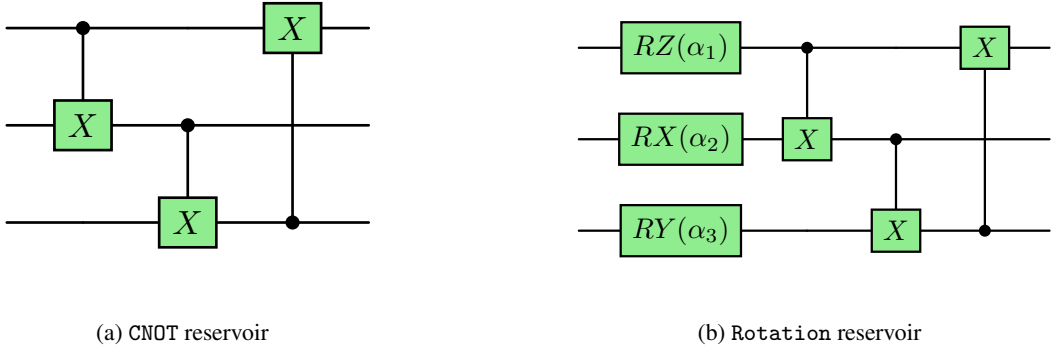
\begin{figure}[!tb]
\centering
\begin{subfigure}[b]{0.42\columnwidth}
\centering
\resizebox{0.75\columnwidth}{!}{ 
\begin{quantikz} [color=black,background color=lightgreen]
&\ctrl{1} & &\gate{X}&\\
&\gate{X}&\ctrl{1}& & \\
&&\gate{X}&\ctrl{-2}& \\
\end{quantikz}
}
\caption{\CNOTreservoir reservoir}
\label{fig:CNOTReservoir}
\end{subfigure}
\hfill
\begin{subfigure}[b]{0.57\columnwidth}
\centering
\resizebox{0.7\columnwidth}{!}{ 
\begin{quantikz} [color=black,background color=lightgreen]
&\gate{RZ(\alpha_1)}& \ctrl{1} & &\gate{X}&\\
&\gate{RX(\alpha_2)}& \gate{X}&\ctrl{1}& & \\
&\gate{RY(\alpha_3)}& &\gate{X}&\ctrl{-2}& \\
\end{quantikz}
}
\caption{\Rotation reservoir}
\label{fig:rotationReservoir}
\end{subfigure}
\caption{Two of the four quantum reservoirs used in \ourApproach}
\label{fig:reservoirs}
\end{figure}

\noindent\textbf{CNOT reservoir (\CNOTreservoir):}
It comprises \CX gates (i.e., \CNOTgate) without randomly assigned parameters. \CX gates are applied similarly to the cyclic entangling structure implemented in the encoder circuits. Fig.~\ref{fig:CNOTReservoir} shows an example of \CNOTreservoir reservoir with 3 qubits.

\noindent\textbf{Harr random reservoir (\Harr):}
It uses a single random operator as the quantum reservoir circuit. The unitary matrix of the operator is sampled from Haar measure~\cite{mezzadri2006generate}.

\noindent\textbf{Ising Mag Traverse Reservoir (\Ising):}
It is one of the widely studied reservoirs. It has a fully connected transverse field, Ising Hamiltonian, in the form of Ising model (a binary quadratic mathematical model). The unitary operator is defined as:
\begin{equation*}
U_{\text{\Ising}} = \exp(-i H \Delta t)
\end{equation*}
where $i$ is the imaginary number, $H$ is the Hamiltonian, and $\Delta t$ is the time step of the Hamiltonian. The Hamiltonian is defined as
\begin{equation}\label{eq:ising}
\textstyle H = \sum_{k,j}{J_{k,j}}{Z_kZ_j} + \sum_{j}{a_jX_j}
\end{equation}
where coefficients $a_j$ and $J_{k,j}$ are the randomly sampled parameters in the unitary matrix, and $X_j$ and $Z_j$ represent the application of the Pauli-X and Pauli-Z gate on the $j$-th qubit.

\noindent\textbf{Rotation reservoir (\Rotation):}
It has single rotation gates and cyclic structure \CX gates. Each qubit has one single-qubit rotation gate randomly selected within \RX, \RY, and \RZ gates. The rotation angles are also randomly assigned. Next, cyclic-structured \CX gates are applied. Fig.~\ref{fig:rotationReservoir} shows a possible circuit with three qubits.

\section{Experiment Design}\label{sec:design}

\noindent{\bf Research questions.}
We investigate the effectiveness of \ourApproach using the following research questions (RQs).\footnote{Confidentiality agreement prevents us from providing the replication package.}
\begin{compactitem}
\item[\textbf{RQ1.}] Which combination of encoder and reservoir of \ourApproach achieves the best prediction performance with a different number of features?

Given that there are two possible encoders and four possible reservoirs, we aim to find their best combination for \ourApproach, that can be used for the experiments in the other RQs.

\item[\textbf{RQ2.}] By using the optimal combination of encoder and reservoir, what is the minimum number of features for which \ourApproach achieves a prediction performance comparable to that achievable using the maximum number of features?

In RQ1, we find the optimal combination of encoder and reservoir for \ourApproach. In this RQ, by using this combination, we aim to find the minimum number of features that can be used by \ourApproach to achieve a prediction performance at least similar to that obtainable by using more features. This RQ helps evaluate the prediction performance of \ourApproach when using a limited number of features. 

\item[\textbf{RQ3.}] How well does \ourApproach perform compared to the baseline when using different numbers of features for predictions? 

This RQ studies whether \ourApproach can help achieve, at least, similar performance with fewer features compared to a classical regression machine learning algorithm.
\end{compactitem}

\smallskip
\noindent{\bf Features and datasets.}
We use the average waiting time (\AWT) as the performance metric, representing the QoS of elevators. For evaluation, Orona provided 12 relevant features for training the model in \ourApproach. These features are:
\begin{inparaenum}[]
\item \feature{1}. Number of upward calls from low-level floors.
\item \feature{2}. Number of upward calls from medium-level floors.
\item \feature{3}. Number of upward calls from high-level floors.
\item \feature{4}. Number of downward calls from low-level floors.
\item \feature{5}. Number of downward calls from medium-level floors.
\item \feature{6}. Number of downward calls from high-level floors.
\item \feature{7}. Average distance of the travel from the upward calls.
\item \feature{8}. Average distance of the travel from the downward calls.
\item \feature{9}. Number of total upward calls in the past 5 minutes.
\item \feature{10}. Number of total downward calls in the past 5 minutes.
\item \feature{11}. Number of calls going upwards
\item \feature{12}. Number of calls going downwards.
\end{inparaenum}

To assess the feasibility of training models with a varied number of features, Orona provided us with specific {\it feature sets} (\featureSetName) built using the 12 features. Table~\ref{tab:feature} reports the details of the feature sets.
\begin{table}[!tb]
\centering
\caption{Feature sets \featureSetName}
\label{tab:feature}
\footnotesize
\begin{tabular}{c|c|l}
\toprule
\multicolumn{2}{c|}{\textbf{Feature set (\featureSetName)}} & \multirow{2}{*}{\textbf{Selected Features}} \\ 
\cmidrule{1-2}
{\bf Number of features} & {\bf ID}\\
\midrule
2 & \featureSet{2} & \feature{11}, \feature{12}\\
\hline
\multirow{2}{*}{3} & \featureSet{3a} & \feature{11}, \feature{12} ,\feature{7} \\
\cline{2-3} 
& \featureSet{3b} & \feature{11}, \feature{12}, \feature{1} \\
\hline
4 & \featureSet{4} & \feature{11}, \feature{12}, \feature{7}, \feature{8} \\
\hline
5 & \featureSet{5} & \feature{11}, \feature{12}, \feature{7}, \feature{8}, \feature{1} \\
\hline
10 & \featureSet{10} & \feature{1}, \feature{2}, \feature{3}, \feature{4}, \feature{5}, \feature{6}, \feature{7}, \feature{8}, \feature{9}, \feature{10} \\
\bottomrule
\end{tabular}
\end{table}
In the table, \featureSet{n} represents a feature set with $n$ features, with $n \in \{2,$ $3,$ $4,$ $5,$ $10\}$; we use $\mathit{3a}$ and $\mathit{3b}$ to distinguish the two sets of 3 features.
Feature sets were built according to these observations.
As elevators of the same product line have different configurations, some of them may lack the ability to record detailed information on elevator calls, such as starting and destination floors. Thus, features \feature{1} to \feature{6} are not in feature sets \featureSet{2}, \featureSet{3a}, \featureSet{4}, and \featureSet{5}. In \featureSet{3b}, instead, only \feature{1} is present. In addition, since \feature{7} and \feature{8} are not available at runtime during operation, these features are not present in \featureSet{2} and \featureSet{3b}. \featureSet{10} is the feature set used to train models of the baseline \mlApproach in~\cite{AitorPaper}.





Orona provided us with 4 datasets of information extracted from the real operation of elevators installed in a 10-floor building: it is the passenger traffic data and the corresponding \AWT values for 4 days. According to the passenger traffic data, we obtained information on the features for a time window of 5 minutes. We used an elevator simulator, i.e., the one provided by \Elevate~\cite{elevateWebsite}, and used in Orona. The simulator was executed over each dataset, and the \AWT in each time window was obtained as the target for training.

\smallskip
\noindent{\bf Independent variables.}
The first independent variable is \texttt{En\-co\-der} type, which takes values in $\{\DHE,$ $\RHE\}$. The second independent variable is {\tt Res\-er\-voir} type, which takes values in $\{\CNOTreservoir,$ $\Harr,$ $\Ising,$ $\Rotation\}$. We investigate which of the 8 combinations of {\tt En\-co\-der} and {\tt Res\-er\-voir} (\combEncRes) gives the best performance for \ourApproach for different feature sets with different number of features (see Table~\ref{tab:feature}), and different datasets. We also compare \ourApproach with the baseline approach \mlApproach, equipped with the two best ML models in~\cite{AitorPaper} (i.e., SVM and regression tree). 

In our study, we use cross-validation, where each dataset was treated as the test dataset while the remaining 3 datasets served as the training dataset. Thus, we conducted four experiments for every feature set \featureSet{n} and combinations of \combEncRes. The experiments were labeled using \expDay{n}, with $n \in \{1,$ $2,$ $3,$ $4\}$, to denote the four datasets provided by Orona.

\smallskip
\noindent{\bf Parameter settings.}
For feature normalization, we used the following range [0, $\pi$]. For other parameters, we used default parameter settings, i.e., encoder depth set to $1$, one qubit for each feature, and depth of \CNOTreservoir and \Rotation reservoirs set to 10. We repeat each experiment 30 times to account for randomness~\cite{arcuri2011practical}. 
%
As parameter values for two versions of the baseline \mlApproach, we use the same used in the original study~\cite{AitorPaper}.

We implemented \ourApproach with Qreservoir package~\cite{qreservoirWebsite} of framework Qulacs~\cite{suzuki2021qulacs}, together with its quantum simulator. Experiments were run on a AMD EPYC 7601 32-core processor.

\smallskip
\noindent{\bf Evaluation metrics and statistical tests.}
For RQ1, for each dataset \expDay{i}, feature set \featureSet{j}, and combination \combEncRes, we assessed the prediction quality using the {\it mean squared error}:
\begin{equation*}
\textstyle \mse = \frac{1}{P}\sum_{j=1}^{P}(\target{j}^{\mathit{pre}} - \target{j})^2
\end{equation*}
where $\target{j}^{\mathit{pre}}$ represents the target value predicted by the model. This indicates the prediction quality of combination \combEncRes for dataset \expDay{i}, using feature set \featureSet{j}. Since we repeated experiments 30 times, we obtained 30 \mse values $\mse_1,$ $\ldots,$ $\mse_{30}$. We computed \textit{average \mse} value as $\amse = \nicefrac{\sum_{i=1}^{30}\mse_i}{30}$.

Then, for each \featureSet{j} in \expDay{i}, we ranked the eight combinations \combEncRes based on \amse values. Finally, for all \featureSet{j} in all datasets, we picked the pair with the highest rank in most cases.

For RQ2, we compared the performance of \ourApproach across all \featureSet{j} in terms of \mse. First, we performed the Kruskal–Wallis test to determine whether overall differences exist for the 30 runs across all the \featureSet{j}. If the test reveals a p-value less than 0.05, it means that there are significant differences among some pairs of \featureSet{j}. In that case, we compared all pairs of \featureSet{j} with the Mann–Whitney U test combined with \Atwelve as the effect size. If the p-value of a pair, e.g., \featureSet{2} and \featureSet{10} is less than 0.05, then it means that significant differences exist between them. Since we are performing multiple comparisons, we also corrected p-values with the Holm–Bonferroni method. Finally, the strength of significance was tested with the \Atwelve statistics. An \Atwelve value lower than 0.5 means that the likelihood that the first pair is better than the second one, e.g., \featureSet{2} is better than \featureSet{10}, and vice versa. According to~\cite{kitchenham2017robust}, the effect size can be interpreted as: \emph{Small} in $(0.34, 0.44]$ and $[0.56, 0.64)$; \emph{Medium} in $(0.29, 0.34]$ and $[0.64, 0.71)$; \emph{Large} in $[0, 0.29]$ and $[0.71, 1]$. 

For RQ3, we compared \ourApproach trained over each feature set \featureSet{i} ($i \in \{2,$ $3a,$ $3b,$ $3,$ $5$, $10\}$), with the two best classical ML algorithms in \mlApproach~\cite{AitorPaper} trained with all the features \featureSet{10}. In this case, we compared 30 \mse values of \ourApproach with one value of \mlApproach. Thus, we performed a one-sample Wilcoxon test to compare a sample with one value. As an effect size measure, we used Cohen's $d$~\cite{cohen1969statistical} and used an existing guide to interpret its magnitude as follows: \emph{Small} if $0 < d < 0.2$ or $-0.2 < d <0$; \emph{Medium} if $0.2 \le d \le 0.8$ or $-0.8 \le d \le -0.2$: \emph{Large} if $d > 0.8$ or $d < -0.8$. If $d$ is lower than $0$, it means that \ourApproach is better than \mlApproach (i.e., lower \mse values). Otherwise, $d$ greater than $0$ means that \ourApproach is worse.

\section{Results and Analyses}\label{sec:results}

\subsection*{RQ1: Assessment of Encoding and Reservoirs}
\begin{figure}[!tb]
\centering
\includegraphics[width=0.6\columnwidth]{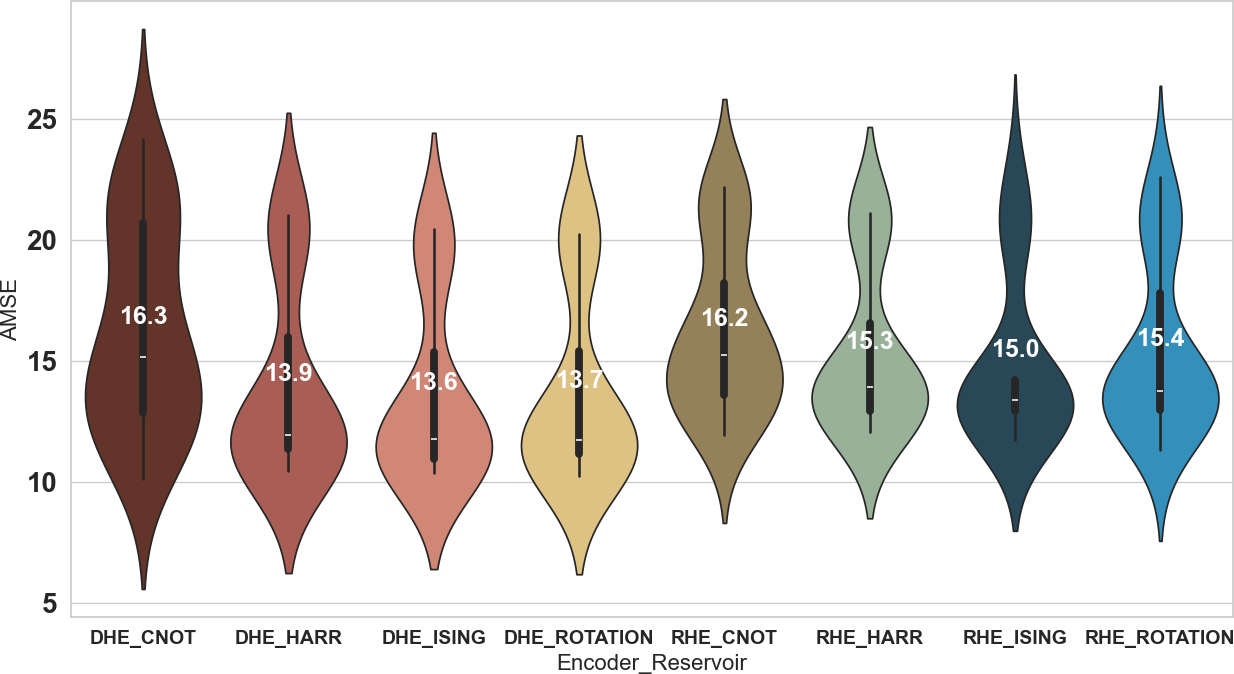}
\caption{RQ1 -- \amse of 8 combinations \combEncRes}
\label{fig:rq1_box}
\end{figure}

First, we evaluate the performance of the 8 combinations of \combination{{\tt Encoder}}{{\tt Re\-se\-rvoir}}. To analyze the overall performance of each combination through various feature sets \featureSetName for all experiments \expDayName, we calculate the \amse of 30 runs for each \emph{setting}, i.e., pair of feature set \featureSetName and dataset \expDayName. Since there are 6 feature sets and 4 datasets, there are 24 settings. Fig.~\ref{fig:rq1_box} reports violin plots where each data point represents the \amse for each setting, and each violin represents the performance of each combination \combEncRes.
%
We observe that most \amse values range from 10 to 25, and that the values of \combination{\DHE}{\Harr}, \combination{\DHE}{\Ising}, and \combination{\DHE}{\Rotation} are the best (i.e., lower than the other ones).

There are also two outlier data points for \combination{\RHE}{\Ising} and one for \combination{\RHE}{\Rotation} (not shown in the figure). We closely examine the \mse values generated by each run with various settings. 
We identify the runs that produced \mse values exceeding 25, resulting in 14 runs. Among them, 11 were generated by the \combination{\RHE}{\Ising} combination, while the remaining three were produced by \combination{\RHE}{\Rotation}. This observation suggests that these two combinations are unstable. As detailed in Sect.~\ref{sec:approach}, the \RHE encoder is randomly generated. Furthermore, the randomly chosen Hamiltonian in the \Ising model and the randomly selected rotation gates (along with their angles) in \Rotation may contribute to such instability. Hence, it is highly likely that these random aspects contributed to a sub-optimal reservoir circuit, resulting in unstable prediction performance.

\begin{figure}[!tb]
\centering
\includegraphics[width=0.6\columnwidth]{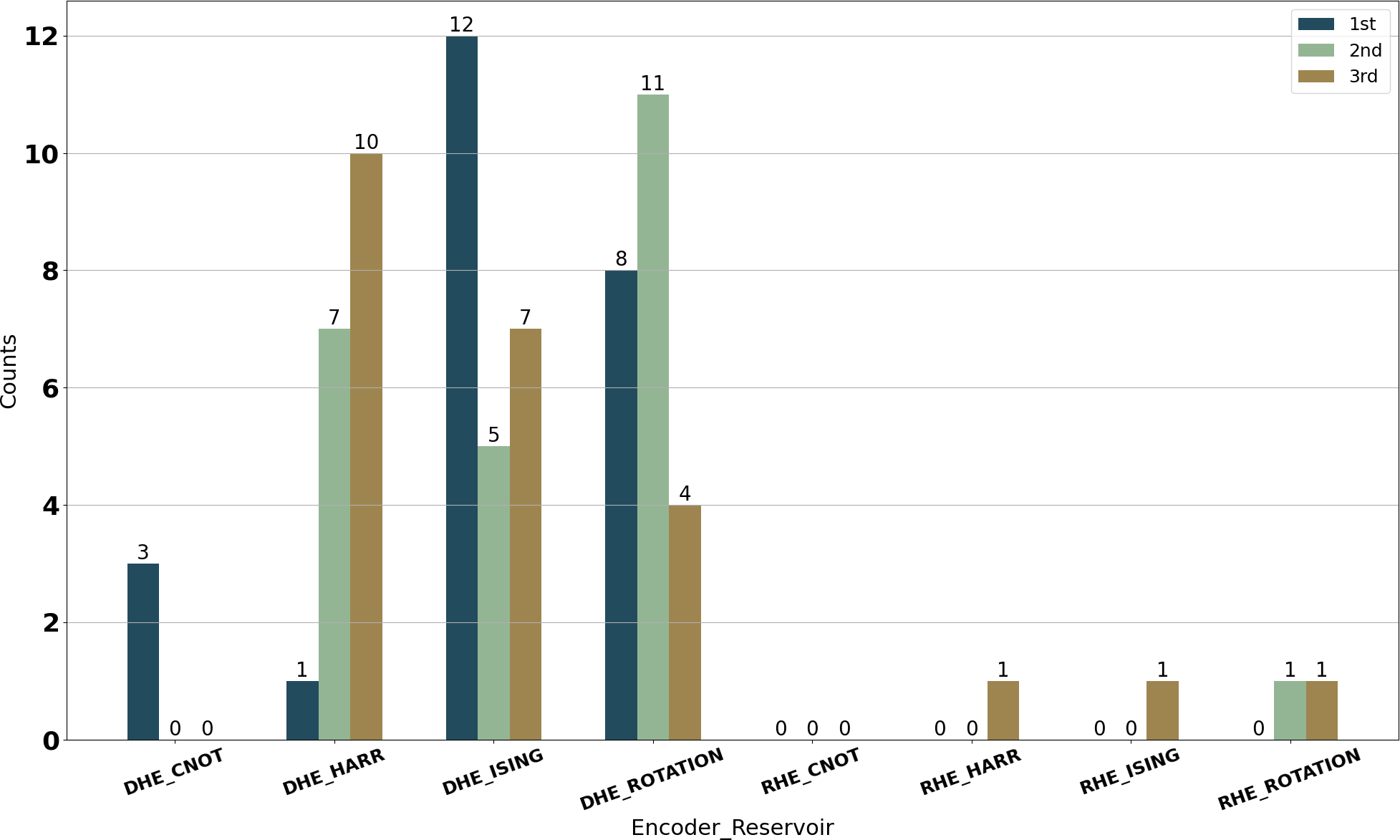}
\caption{RQ1 -- 1st, 2nd, and 3rd position for each combination of \combEncRes}
\label{fig:rq1_ranking}
\end{figure}

To analyze the performance of the combinations in detail, we rank the \amse of each combination to obtain the top three combinations for each setting. We report the results in Fig.~\ref{fig:rq1_ranking}.
We notice that in most settings, \DHE encoder is in the 1st, 2nd, and 3rd combination. \combination{\DHE}{\Harr}, \combination{\DHE}{\Ising}, and \combination{\DHE}{\Rotation} ranked top 3 in the majority of settings. Finally, we observe that \combination{\DHE}{\Ising} ranks the first in 12 out of 24 settings.

\begin{tcolorbox}[size=title, colframe=green!10, width=1\linewidth, colback=green!10, breakable]
\textbf{Conclusions for RQ1:}
Overall, the \Ising reservoir combined with the \DHE encoder enables \ourApproach to perform the best. Hence, we suggest using this combination in \ourApproach.
\end{tcolorbox}


\subsection*{RQ2: \ourApproach's Performance with Different Feature Sets}\label{subsec:RQ2}
\begin{figure}[!tb]
\centering
\begin{subfigure}[b]{0.495\columnwidth}
\centering
\includegraphics[width=0.95\linewidth]{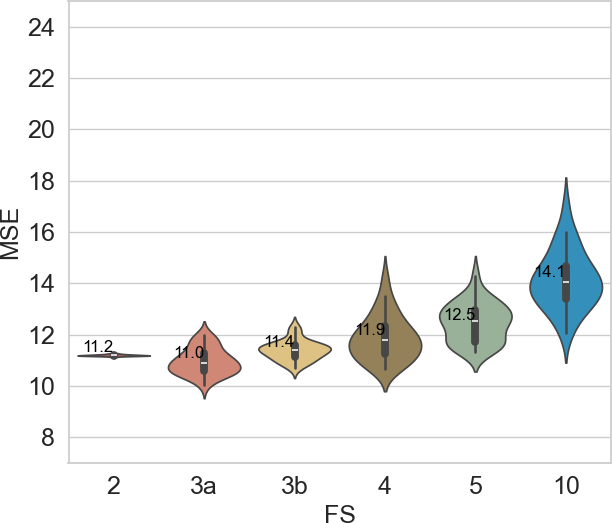}
\caption{\expDay{1}}
\label{fig:rq2_box_day1}
\end{subfigure}
\begin{subfigure}[b]{0.495\columnwidth}
\centering
\includegraphics[width=0.95\linewidth]{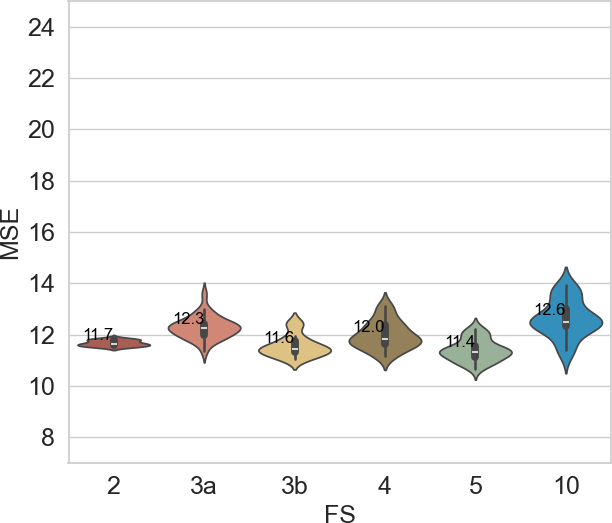}
\caption{\expDay{2}}
\label{fig:rq2_box_day2}
\end{subfigure}

\begin{subfigure}[b]{0.495\columnwidth}
\centering
\includegraphics[width=0.95\linewidth]{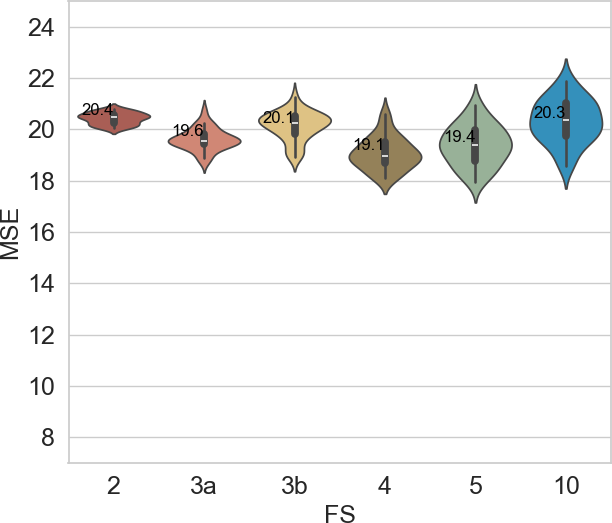}
\caption{\expDay{3}}
\label{fig:rq2_box_day3}
\end{subfigure}
\begin{subfigure}[b]{0.495\columnwidth}
\centering
\includegraphics[width=0.95\linewidth]{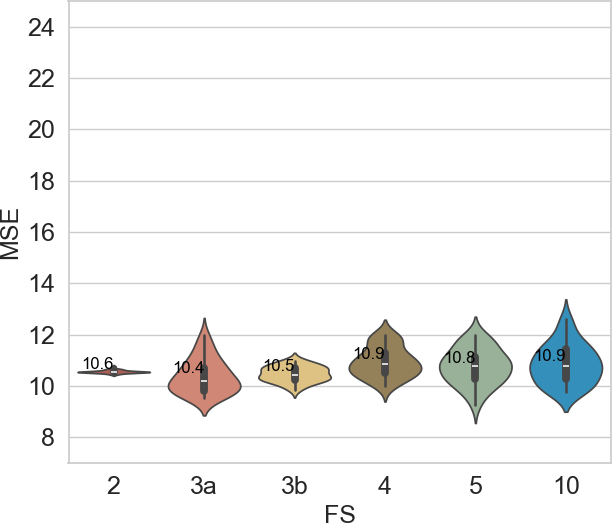}
\caption{\expDay{4}}
\label{fig:rq2_box_day4}
\end{subfigure}
\caption{RQ2 -- \mse of \ourApproach with different \featureSetName}
\label{fig:rq2_box}
\end{figure}

We choose the best combination of \combEncRes in RQ1, i.e., \textit{\DHE\_\Ising}, to compare the performance of \ourApproach with different feature sets \featureSetName of different sizes. For each dataset, we draw the violin plot of \mse values shown in Fig.~\ref{fig:rq2_box}.
%
We notice that the fluctuation in \mse values produced by \ourApproach decreases with smaller feature sets, as shown by the low variance of \featureSet{2} in all four datasets. The possible reason is that, since the \DHE encoder remains constant for a specific feature set, the randomness is only introduced by the quantum reservoir, specifically by the unitary matrix determined by the \Ising Hamiltonian. Because fewer features require few qubits, according to Eq.~\ref{eq:ising}, there is a reduced need for parameters to be randomly selected.

We also observe that \ourApproach with less than 10 features can consistently achieve performance comparable to \ourApproach with 10 features (i.e., similar or even lower \mse values). Since 10 features require 10 qubits, this increases the likelihood that random parameters are assigned to the \Ising Hamiltonian; this can generate a suboptimal reservoir circuit, which may exhibit limited dynamics, potentially leading to a less effective mapping from the classical data into the observed values. Moreover, since we aim to use the most efficient quantum circuit, the encoder depth qubit is 1. However, in the case of complex quantum circuits involving 10 qubits, a higher depth may demonstrate increased effectiveness.

For \expDay{3}, the \mse values are generally higher than in other experiments. The cause of this is that the testing dataset includes an outlier with an \AWT value exceeding 35 seconds, whereas the \AWT values in the training dataset fall within the range of 0 to 26 seconds, which results in difficulty for \ourApproach to predict that value. 

\begin{figure}[!tb]
\centering
\begin{subfigure}[b]{0.495\columnwidth}
\centering
\includegraphics[width=0.95\linewidth]{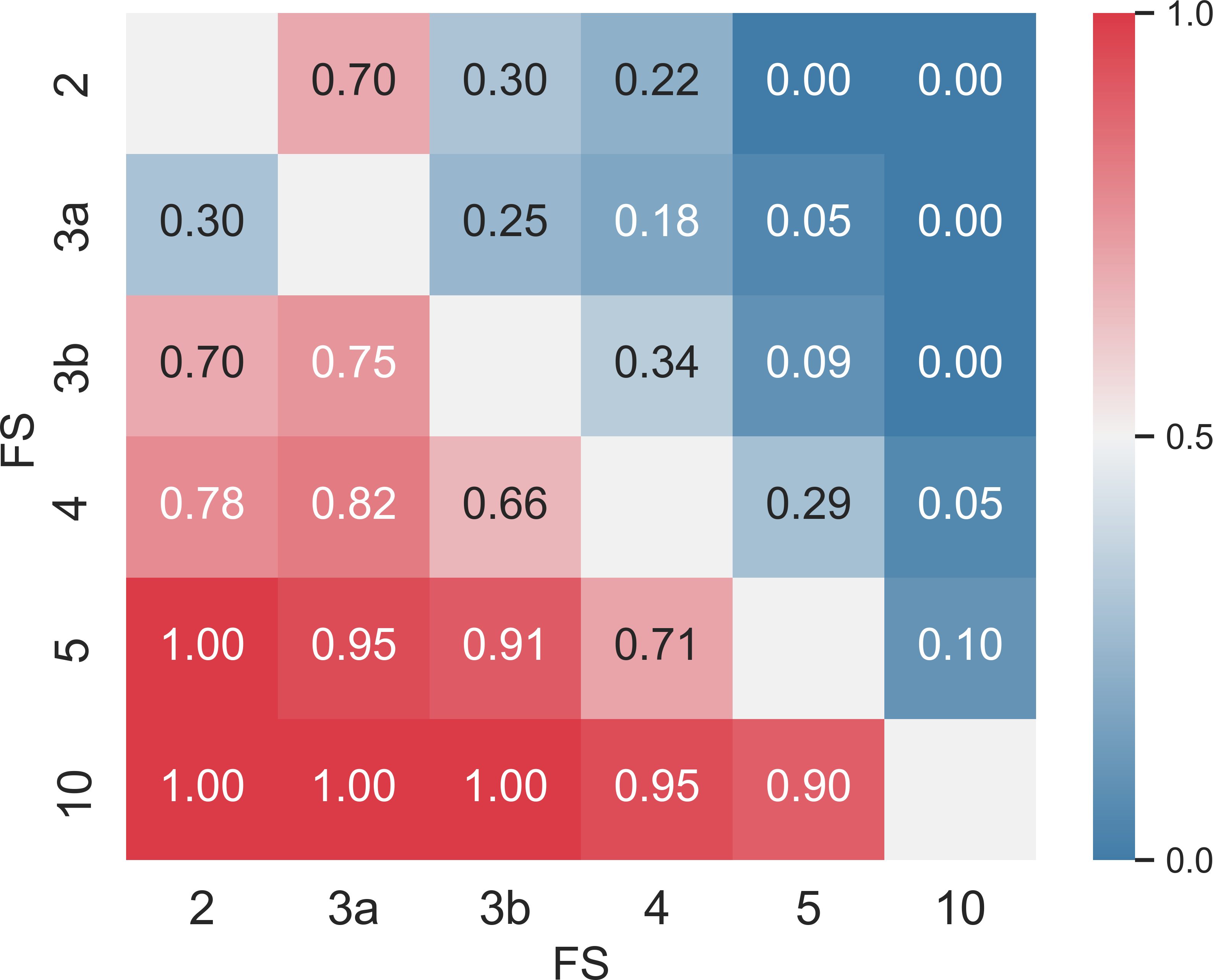}
\caption{\expDay{1}}
\label{fig:rq2_heat_day1}
\end{subfigure}
\begin{subfigure}[b]{0.495\columnwidth}
\centering
\includegraphics[width=0.95\linewidth]{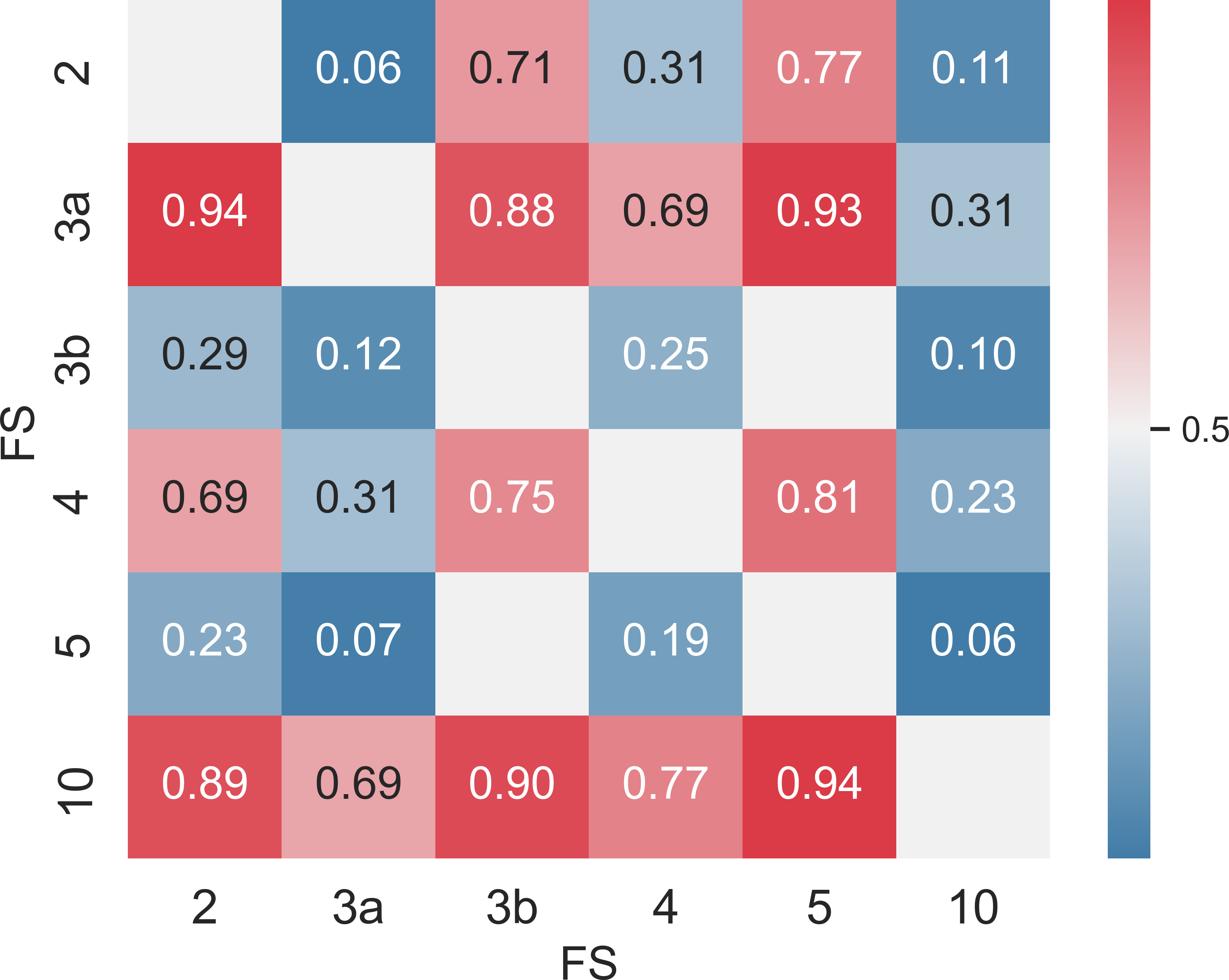}
\caption{\expDay{2}}
\label{fig:rq2_heat_day2}
\end{subfigure}

\begin{subfigure}[b]{0.495\columnwidth}
\centering
\includegraphics[width=0.95\linewidth]{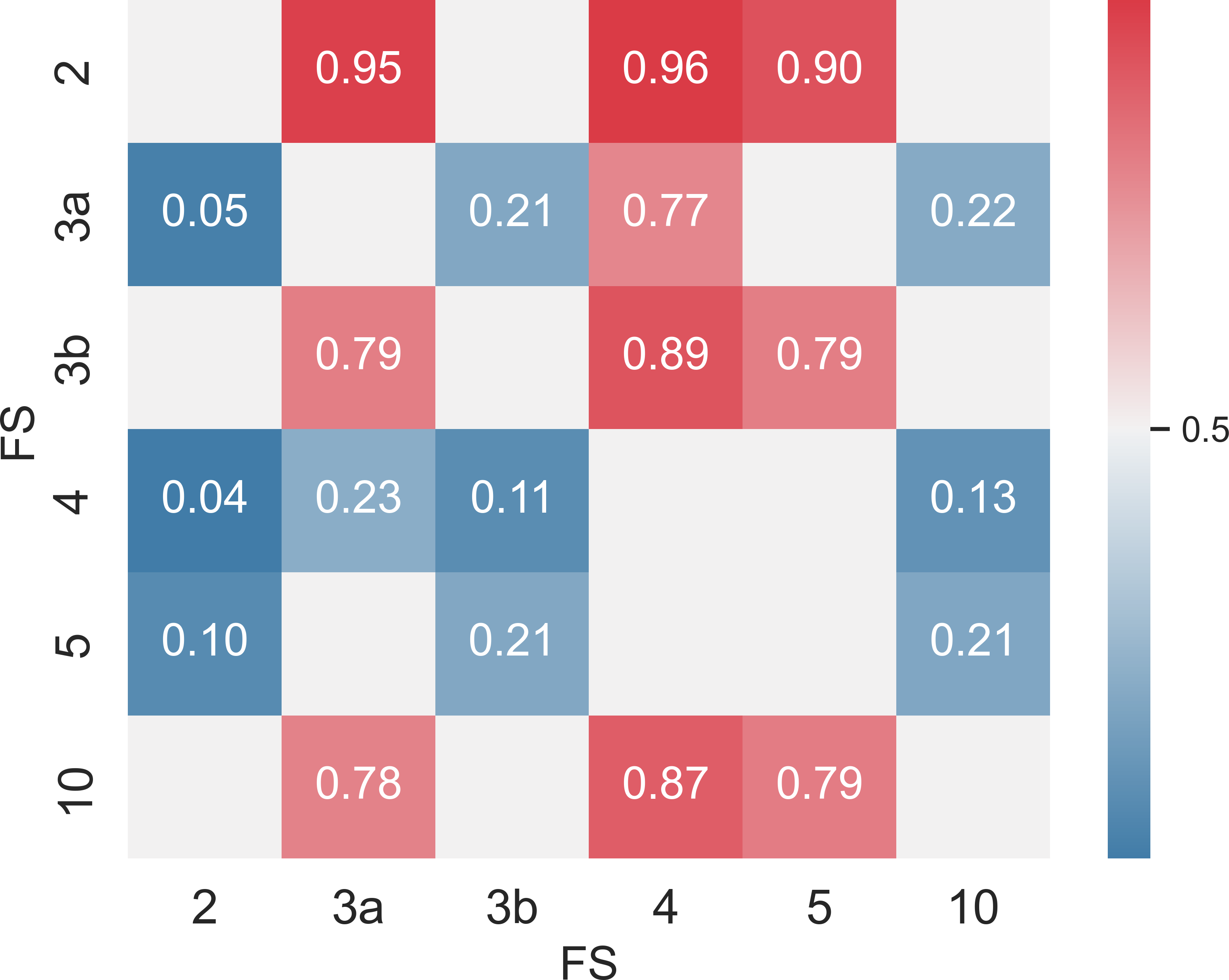}
\caption{\expDay{3}}
\label{fig:rq2_heat_day3}
\end{subfigure}
\begin{subfigure}[b]{0.495\columnwidth}
\centering
\includegraphics[width=0.95\linewidth]{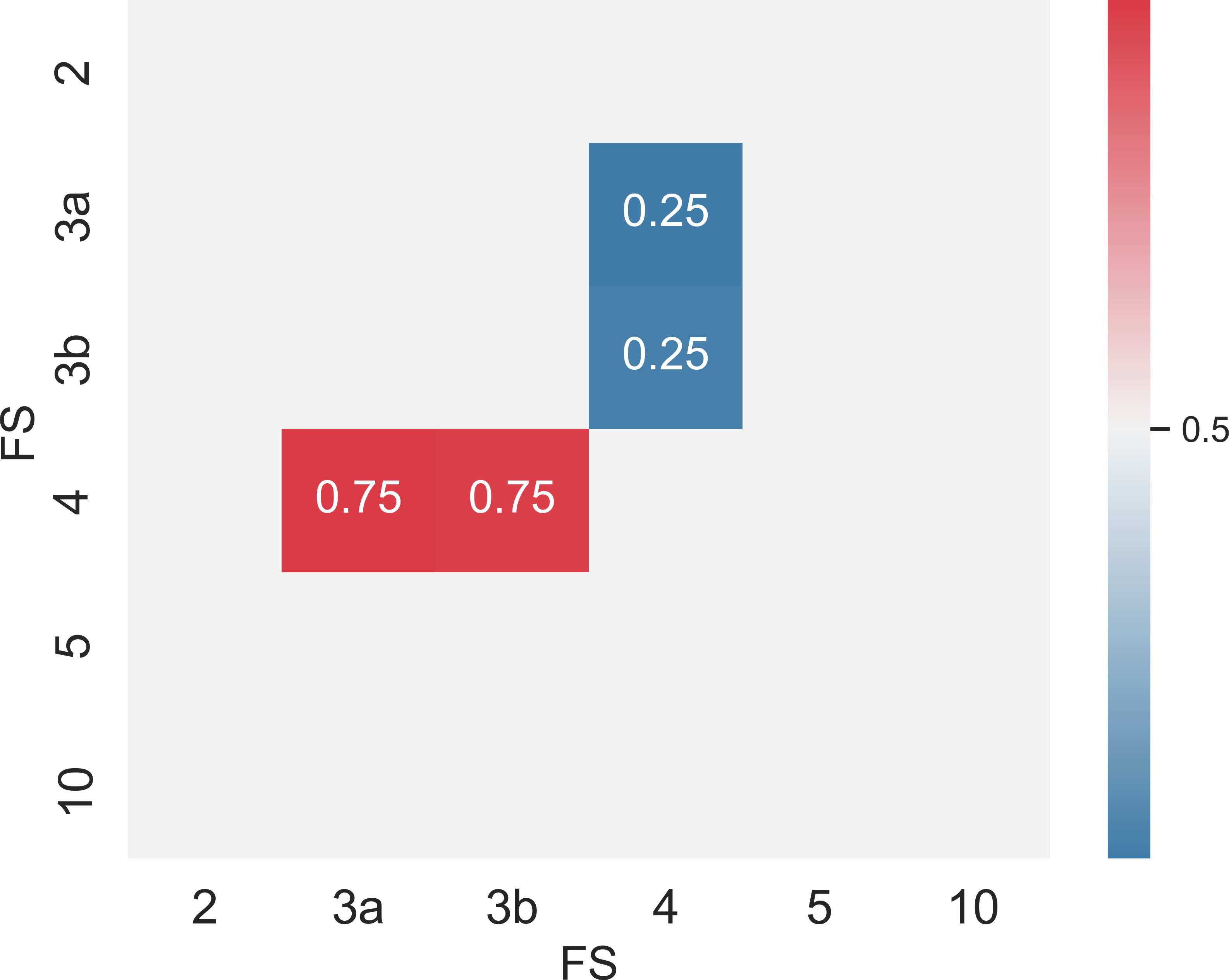}
\caption{\expDay{4}}
\label{fig:rq2_heat_day4}
\end{subfigure}
\caption{RQ2 -- Comparison of \ourApproach with different \featureSetName (\Atwelve values of the comparison of the configuration on the y-axis vs. the configuration of the x-axis)}
\label{fig:rq2_heat}
\end{figure}

We have also used statistical tests to analyze the comparison among those configurations in detail. First, for each dataset, we applied the Kruskal-Wallis test among the different versions of \ourApproach trained with different feature sets \featureSetName. We found that all p-values are lower than 0.05, meaning there are significant differences among \mse values of different \featureSetName. Then, we compared the \mse values of each pair of configurations using the Mann–Whitney U test, together with Holm-Bonferroni correction according to Sect.~\ref{sec:design}. In Fig.~\ref{fig:rq2_heat}, gray cells indicate no significant difference between the corresponding two configurations.
We also report the \Atwelve values for the other cells. Blue color (corresponding to \Atwelve lower than 0.5) means that the configuration on the y-axis is better (i.e., lower \mse) than the configuration on the x-axis; red color means the opposite. In \expDay{1}, \ourApproach with \featureSet{3a} is the most competitive. In \expDay{2} and \expDay{3}, no configuration achieves significantly better performance than \ourApproach with \featureSet{5}. In \expDay{3}, \ourApproach with \featureSet{4} is the most competitive configuration. However, in \expDay{4}, there is no significant difference among most configurations. Moreover, \ourApproach with \featureSet{10} is significantly worse than any other \featureSetName in \expDay{1} and \expDay{2}. In all four experiments, \ourApproach with \featureSet{10} is not significantly better than any other configuration, which aligns with the previous finding that it requires further setting.

\begin{tcolorbox}[size=title, colframe=green!10, width=1\linewidth, colback=green!10, breakable]
\textbf{Conclusions for RQ2:} Overall, \ourApproach with few features outperforms \ourApproach with the maximum number of features 10. This shows the effectiveness of QELM in our industrial context, where few features are available for real-time predictions.
\end{tcolorbox}

\subsection*{RQ3: Comparison of \ourApproach with \mlApproach}\label{subsec:RQ3}
We compare the \mse values of all the versions of \ourApproach (i.e., using all the feature sets \featureSetName) with those of the baseline \mlApproach using the feature set with 10 features, i.e., \featureSet{10}. For \mlApproach, we consider its two best versions in~\cite{AitorPaper}, i.e., when using SVM and regression tree. We identify the two versions as \mlApproachSVM and \mlApproachRT. Since the \mse values of \mlApproachSVM and \mlApproachRT are fixed for the same training and testing datasets, we perform a one-sample Wilcoxon signed rank test to compare \mse values of \ourApproach with various \featureSetName with that of \mlApproachSVM and \mlApproachRT in each experiment \expDayName. The results show that all p-values are less than 0.05, which indicates the significant difference between \ourApproach and the two classical algorithms of \mlApproachSVM and \mlApproachRT.

Next, we compute Cohen's $d$ effect size to see the magnitude of the differences. The results show that all calculated $d$ values are lower than $-1$, which indicates that all \mse values generated by our approach are greatly smaller than those generated by \mlApproachSVM and \mlApproachRT in each experiment. As a result, we conclude that \ourApproach with any number of features performs significantly better than the state of the practice classical machine learning algorithms.

For each experiment, we also counted the number of runs that the \mse values generated by \ourApproach with different \featureSetName are larger than that of \mlApproachSVM and \mlApproachRT. For \expDay{1} and \expDay{2}, all \mse values generated by \ourApproach with all feature sets, are smaller than those of both \mlApproachSVM and \mlApproachRT. For \expDay{3} with \featureSet{10}, there are two runs in which the generated \mse values of \ourApproach are larger than the ones generated by \mlApproachSVM. As in RQ2, \ourApproach with \featureSet{10} is the worst configuration for this dataset.
In \expDay{4}, for \ourApproach with \featureSet{3a}, \featureSet{4}, \featureSet{5}, \featureSet{10}, there are 1, 6, 3, and 4 out of 30 runs in which \ourApproach generated \mse values higher than the ones generated by \mlApproachRT. 

These results align with RQ2's results. For example, in \expDay{3}, we can see that \ourApproach with \featureSet{10} generated the highest \mse value according to the violin plot in Fig.~\ref{fig:rq2_box_day3}, which provides the explanation of the two runs in which it is worst than \mlApproachSVM. Moreover, as shown in Fig.~\ref{fig:rq2_heat_day4}, in \expDay{4}, in general, \ourApproach with all feature sets has similar performance, except for \ourApproach with \featureSet{4}, which is significantly worse than \featureSet{3a} and \featureSet{3b}. This observation explains why \ourApproach with \featureSet{4} contains the highest number of runs (6) in which \mse values of \ourApproach are higher than those of \mlApproachRT.

In any case, in most of the cases, \ourApproach performs better than the baselines (i.e., it has lower \mse values).

\begin{tcolorbox}[size=title, colframe=green!10, width=1\linewidth, colback=green!10, breakable]
\textbf{Conclusions for RQ3:}
For the same prediction task in our industrial context, \ourApproach outperforms classical machine learning approaches (i.e., \mlApproachSVM and \mlApproachSVM), by using less features. This demonstrates the potential of QELM.
\end{tcolorbox}

\section{Discussion}\label{sec:discussion}
We present a discussion and key lessons learned in this section that are valuable for practitioners and researchers.

\subsection{Improvement Over State of the Practice} \label{subsec:impro}
The comparison with the state of the practice approach \mlApproach demonstrated that QELM, indeed, helps our approach perform significantly better than classical machine learning algorithms, thus proving evidence of improvement. We further compared time cost associated with \ourApproach and classical machine learning algorithms. The training time, even on the simulator, was around 1 second or even lower, whereas the inference time was much lower than 1 second for the four datasets we experimented with. Such time cost is practically negligible in our context and, so, shows that even with a negligible cost \ourApproach can perform better than the state of the practice approach. Notice that executing QELM on a real quantum computer will be much faster than the simulators, thereby bringing down these costs even further.

\subsection{Potential Applications in Orona}\label{subsec:Oronaapp}
\ourApproach demonstrated a good performance at predicting \AWT over 5-minute time windows with few features. As discussed before, Orona's main application will be runtime monitoring. Orona continuously maintains its code and deploys changes in operation. This code may have bugs that lead to sub-optimal elevator scheduling, incrementing passenger waiting times and so reducing comfort. \ourApproach will help Orona effectively and efficiently predict waiting times at runtime, serving as a runtime monitor to take corrective actions (e.g., roll-back to a previous version) when an issue is found.

Besides this, \ourApproach can have other applications in Orona. For instance, it can be used for regression testing at the HiL test level when not all features can be obtained. The classical machine learning algorithms employed in the state-of-the-practice approach~\cite{AitorPaper} could be replaced with QELM for this purpose.

Another possible application is the deployment of \ourApproach in digital twins (DTs) of elevators. In our recent work with Orona~\cite{DTEvolution}, we built DTs of elevators with classical machine learning algorithms. Also, we supported the evolution of DTs in response to the evolution of elevators. These DTs use neural network models for waiting time predictions, which is a task similar to that of this work. This prediction task could also benefit from QELM.

\subsection {Generalizability to Other Practical Contexts } 
We demonstrated the application of QELM in one industrial context for one prediction task. However, it is important to note that our implementation of QELM is not specific to this industrial context, and it can also be applied to other prediction tasks in other industrial contexts with different feature sets. Whether the current best combination of encoding type and reservoir types will work for other contexts requires further investigation. However, we find that the current demonstration of QELM in our industrial context is representative, demonstrating the benefits of QELM. Our work could inspire practitioners and researchers to experiment with QELM and other quantum machine learning techniques to overcome software engineering problems that require regression problems and the number of features to be used is limited. 

\subsection{Research Implications} \label{subsec:research}
We discuss research implications for classical and quantum software engineering, and for theoretical foundations of QELM.

\smallskip
\noindent\textbf{Classical and Quantum Software Engineering.}
Our work opens up two new research directions within the software engineering area (both classical and quantum). \textbf{First}, it opens up areas in applied quantum machine learning (QML) in software engineering applications. In particular, QELM and other new QML algorithms can be studied for various classical software engineering problems, such as regression testing, defect prediction, software development cost prediction, and software performance predictions. Similarly, even though quantum software engineering (QSE) methods~\cite{zhao2020quantum,AliCACM2022} are limited, QML methods can be applied to solve QSE problems such as quantum software testing and debugging. Since quantum software operates on quantum states, there might not be a need for quantum encoders. \textbf{Second}, we also see the opportunities for empirical research focusing on empirically evaluating various parameters of QELM and other QLM algorithms. For instance, for QELM, we used default settings for parameters such as encoder depth equal to 1 and one feature mapped to exactly one qubit. There is a need for more extensive empirical evaluations to study how varying these parameters affect the prediction performance; hence, more carefully planned empirical studies are required. Such empirical evaluations, for example, will provide evidence about the performance of various parameter configurations for different problems. Such evidence will also result in developing guides (currently missing) for conducting empirical evaluations for QELMs, guiding users, for example, about which settings to use, which statistical tests to perform, and which cost-effectiveness criteria to use.

\smallskip
\noindent\textbf{Theoretical Foundations in QELM.} This paper applied existing QELM implementations (including encoding and reservoir types) to our industrial problem. Such QELM implementation is preliminary, and more research is needed to design better encoders and reservoirs. One potential direction to investigate is to design domain-specific encoders and reservoirs that work for specific domains of problems (e.g., elevator operation optimization). For further discussion on the research opportunities offered by QELM, we refer interested readers to~\cite{QELMOpportun}.

\subsection{Limitations} \label{subsec:limitation}
Our application has some limitations, given that QELM is a very new area. First, we executed QELM on a quantum computer simulator with Qulacs framework. We used an ideal simulator, i.e., with no hardware noise. However, currently, quantum computers have hardware noise that affects the computations they perform. As a result, if we run QELM on a real quantum computer, our results will be affected. However, currently, it is not possible to directly execute QELM on real quantum computers due to practical hardware constraints. Moreover, we would like to highlight that QELM is one of the algorithms inherently designed to deal with hardware noise; thus, we expect them to perform well even on real quantum computers. On the other hand, note that as the number of qubits increases, the simulation time on quantum computer simulators increases exponentially, thereby limiting the practical application of QELM. Nonetheless, once QELM can be executed on quantum computers, the execution time, even with a higher number of qubits, will be much lower than execution on simulators.

\section{Threats to Validity} \label{sec:threats}

We normalize the features in the range of [0, $\pi$]. Different normalization ranges may affect the results, thereby requiring additional investigation. QELM's parameters were set to default values, i.e., encoder depth to 1, one qubit mapped to one feature, and depths of quantum reservoirs to 10. Such default values have shown sufficiently good results in the previous studies~\cite{ghosh2021realising}. Finally, to select the best combination of encoding and reservoir type for \ourApproach, we performed a systematic experiment as reported in RQ1. 

For \mlApproach used as the baseline, we picked its settings from the original work~\cite{AitorPaper}, i.e., the maximal number of decision splits per tree of 25 for the regression tree, whereas the Gaussian kernel function was used. Other parameters were set to default settings. For comparing \ourApproach with the baselines, we picked the two best machine learning algorithms from the original work~\cite{AitorPaper}. 

There is inherent randomness in QELM; thus, we repeat each testing task 30 times to account for it, as common practice in the evaluation of randomized algorithms~\cite{arcuri2011practical}. We collected the results of the 30 repetitions and applied appropriate statistical tests based on established guides from the literature on randomized algorithms~\cite{arcuri2011practical}. 

We experimented with only four datasets corresponding to four days of elevator operations in one building. Naturally, it poses threats to the generalizability of our results since data for different building configurations, different days, and different configurations of the dispatching algorithm are possible. This dataset, however, was obtained from a real building, being a good representation of the real-world. The access to this data is difficult even for Orona's engineers as it requires significant resources. 
Once we have access to more data, we can easily extend the empirical evaluation.


\section{Related Works}\label{sec:related}
Multiple recent studies have focused on enhancing the software quality of elevator dispatching algorithms from different perspectives, including, robustness analysis~\cite{han2023uncertainty,han2022elevator}, testing~\cite{AitorPaper,ayerdi2022performance,ayerdi2021generating,arrieta2022automating,galarraga2021genetic}, debugging~\cite{ValleISSTA} and repair~\cite{ValleICSE}. Our work is close to those from Ayerdi et al.~\cite{ayerdi2021generating,ayerdi2022performance}, since, given a set of passengers, predicting which the output should be is not trivial. There are two key differences between this work and the ones from Ayerdi et al.~\cite{ayerdi2021generating,ayerdi2022performance}. Firstly, our approach aims to predict the waiting time of the system by applying QELM, whereas their approach leverages metamorphic testing. Secondly, our approach is applicable both at design-time as well as operation-time, whereas their technique is solely focused on operation-time. Similar to our approach, others aim at predicting waiting times of passengers~\cite{AitorPaper,arrieta2021using}. Nevertheless, as explained in previous sections, their approach has certain limitations that we overcome in this paper by using QELM instead of classical ML-based techniques. Specifically, QELM can provide advantages when not all features are available, which is the case for runtime monitoring of these systems when deployed in operation. In summary, to the best of our knowledge, this is the first paper that applies QELM to predict QoS metrics of a system of elevators.

In the past, several studies tackled the test oracle problem by employing machine learning by following different strategies (e.g., automatically predict the test verdict~\cite{braga2018machine,gholami2018classifier,makondo2016exploratory,shahamiri2010automated}). Similar to our approach, several studies aim at predicting which is the expected output~\cite{aggarwal2004neural,ding2016machine,jin2008artificial,monsefi2019performing,sangwan2011radial,shahamiri2011automated,shahamiri2012artificial,singhal2016approach,vanmali2002using,singhal2014generation,mao2006neural,zhang2019automatic}. There are significant differences between our approach and these ones. The first one is that we are applying QELM instead of traditional machine-learning techniques (e.g., SVMs), with the goal of reducing the feature space for our technique to be applicable at the operation time. The second one is that we apply these techniques in the context of Cyber-Physical Systems (CPSs), unlike most of the aforementioned approaches, which focus on testing software code. Lastly, unlike the previously mentioned approaches, we apply our approach in the context of a real industrial case study system. In the context of CPSs, and more particularly autonomous driving systems, several approaches have targeted runtime monitoring by leveraging advanced techniques to measure the uncertainty of the DNNs taking control of the vehicle~\cite{stocco2020misbehaviour,stocco2022thirdeye}. Unlike our industrial system, such approaches are applicable for cases where the system's control is driven by ML-based techniques, which is not the case in our industrial system. 

Quantum reservoir computing (QRC) and QELM offer ease in model training with quantum reservoir circuits for linear regression models for machine learning tasks (e.g., prediction)~\cite{QELMOpportun,fujii2021quantum,tanaka2019recent}. The difference between QRC and QELM is that tasks accomplished by QRC employs the memory of the reservoir, while QELM uses memoryless reservoirs, which makes the training of QELM easier~\cite{QELMOpportun}. Given that these techniques can potentially be executed on quantum computers with high computational power complemented with the increased availability of quantum computing platforms, there has been increased interest in using QRC or QELM for solving problems in the classical domain~\cite{fujii2017harnessing,nakajima2019boosting,martinez2020information,nokkala2021gaussian,PhysRevResearch.3.013077,xia2023configured} and quantum domain~\cite{domingo2023taking,domingo2022optimal,kawai2020predicting,ghosh2021realising, innocenti2023potential}. However, these works focus on benchmark problems. In contrast, \ourApproach provides implementation of QELM for solving a practical industrial problem with real operational data from industrial elevators.

\section{Conclusion}\label{sec:conclusion}
We presented an industrial application of quantum extreme learning machine (QELM) for solving the waiting time prediction task in the context of elevators. By using four real datasets from the elevators' real operation, we demonstrated that QELM could offer benefits by performing significantly better prediction even with a fewer number of features than the classical ML algorithms that rely on a higher number of features. Our results demonstrate the potential of using QELM to solve various software engineering problems requiring machine learning algorithms. In the future, we would like to experiment with our approach with more datasets and other application contexts (i.e., integration with digital twins of elevators and real-time prediction) in the context of Orona. Moreover, we will apply our approach to other industrial and real-world applications. Given that we ran our approach on quantum computer simulators with no noise, we will include noise on the encoder and reservoir to study how it affects our results. On the foundational side, we would also like to design new domain-specific encoders and quantum reservoirs for different applications.

\section*{Acknowledgments}
Xinyi Wang is supported by Simula's internal strategic project on Quantum Software Engineering and NordIQuEst project funded by NordForsk. Shaukat Ali is supported by Qu-Test (Project\#299827) funded by the Research Council of Norway and Simula's internal strategic project on Quantum Software Engineering. Aitor Arrieta is part of the Software and Systems Engineering research group of Mondragon Unibertsitatea (IT1519-22), supported by the Department of Education, Universities and Research of the Basque Country.
Paolo Arcaini is supported by Engineerable AI Techniques for Practical Applications of High-Quality Machine Learning-based Systems Project (Grant Number JPMJMI20B8), JST-Mirai; and by ERATO HASUO Metamathematics for Systems Design Project (No. JPMJER1603), JST, Funding Reference number: 10.13039/501100009024 ERATO.


\bibliographystyle{abbrv}  
\bibliography{biblio}

\end{document}